\documentclass[preprint,review,12pt,sort&compress]{elsarticle} 

\journal{Annals of Physics}

\usepackage{amsmath}
\usepackage{bm}
\usepackage{xspace}

\usepackage{amsmath}
\usepackage{mathtools}
\usepackage{mathrsfs}
\usepackage{amssymb}
\usepackage{dsfont}
\usepackage{slashed}
\usepackage{empheq}
\usepackage{relsize}
\usepackage{nicefrac}
\usepackage{enumerate}

\usepackage{braket}
\usepackage{amsmath}
\usepackage{mathtools}
\usepackage{mathrsfs}
\usepackage{amssymb}
\usepackage{dsfont}
\usepackage{slashed}
\usepackage{empheq}
\usepackage{relsize}
\usepackage{nicefrac}

\begin{document}

\begin{frontmatter}



\title{Equilibration in fermionic systems}


\author{T.~Bartsch}
\author{G.~Wolschin\corref{cor}}
\ead{g.wolschin@thphys.uni-heidelberg.de}

\address{Institut f{\"ur} Theoretische Physik der Universit{\"a}t Heidelberg, Philosophenweg 16, D-69120 Heidelberg, Germany, EU}

\cortext[cor]{Corresponding author}

\begin{abstract}
The time evolution of a finite fermion system towards local statistical equilibrium is investigated
using analytical solutions of a nonlinear 
partial differential 
equation that had been derived earlier
from the Boltzmann collision term. The solutions
of this fermionic diffusion equation are rederived in closed form, evaluated exactly for simplified initial conditions, and applied to hadron systems at low energies in the MeV-range,
as well as to quark systems at relativistic energies in the TeV-range where antiparticle production is abundant.
Conservation laws for particle number including created antiparticles, and for the energy are
discussed.
\end{abstract}


\end{frontmatter}
\newpage

\section{Introduction}
The evolution of a physical system towards local statistical equilibrium is of general interest. In the realm of quantum physics, fermions and bosons with their respective quantum-statistical properties must be considered separately. Due to the antisymmetry of their states, fermions obey Pauli's principle. Changes of the occupation probabilities of single-particle states are correspondingly suppressed, causing significantly larger local equilibration times as in the case of bosons. 
The latter are not only free to occupy any state, but can also form a condensate at energy $\epsilon=0$, which can have a significant effect on the statistical equilibration process.

In this work we concentrate on the thermalization of fermionic systems, with possible applications to heavy-ion collisions at low (MeV-range), intermediate, and relativistic (TeV-range) energies. At low energies, Pauli's principle suppresses nucleon-nucleon collisions leading into occupied states. Hence, the nucleons (fermions) have a long mean free path and can be considered to move in a self-consistent, time-dependent mean field, giving rise to the time-dependent Hartree-Fock approximation (TDHF) \cite{bkn76}. At intermediate energies up to 
$\sqrt{s_\text{NN}}\simeq 100$ MeV, mean-field effects as well as two-body collisions due to the residual interaction must be considered \cite{wong78,os79,ay80,gww81,og89}. At relativistic energies in the GeV and TeV region, mean-field effects can be neglected. The fermionic Boltzmann collision term is then relevant for the local statistical equilibration of quarks, and the corresponding bosonic collision term for 
 the equilibration of gluons.

The full many-body problem with mean-field and collision term can only be solved numerically. Approximate solutions may be obtained with simplified forms of the collision term. As an example, a phenomenological collision term based on a linear relaxation ansatz that governs the equilibration towards the local mean momentum had been added
to the Wigner transform of the one-body density in nonrelativistic calculations \cite{grww81}. A similar ansatz has also been used in relativistic calculations \cite{bay84,jpbli18}. Obviously it does not properly account for the system's nonlinearity that is imposed by the dynamical effect of two-body collisions. It was therefore proposed by one of us in Ref.\,\cite{gw82}, and in Ref.\,\cite{gw18} for bosons, to replace the relaxation ansatz for the collision term by a nonlinear diffusion equation in momentum space which can be solved analytically. 

In this work, we rederive the exact solutions of the nonlinear fermionic diffusion equation using a different method that yields the same result. We give the explicit form of the solutions for schematic initial conditions, and apply the model to nonrelativistic, as well as to relativistic energies where the creation of particle-antiparticle pairs is included. 

In the next section, we reconsider the derivation of the nonlinear fermionic diffusion equation from the Boltzmann collision term.
Its solution is rederived in Section\,3 using a nonlinear transformation that leads to an exactly solvable Fokker-Planck equation, and compared to the linear relaxation ansatz. It is shown that the analytical solution is identical to the one obtained in Ref.\,\cite{gw82}, where a different solution method was used. In Section\,4, the solution is applied to the thermalization of a hadronic system through the collision term at MeV energies, and a quark system at TeV energies. The conservation laws for particle number and energy are discussed in Section\,5, and the conclusions are drawn in the final section.

\section{Derivation of the nonlinear diffusion equation}

The time-dependent mean-field or TDHF approximation \cite{bkn76} for heavy-ion collisions  is expected to be good for the low-energy domain, where the mean free path of the particles is comparatively long. However, at higher energies the effects of two-body collisions cannot be neglected, so that the mean field description has to be extended in order to include residual interactions \cite{wong78,os79,ay80,gww81,og89}. In particular, in Ref.\,\cite{gww81} a random-matrix model had been used to obtain an equation for the reduced single-particle density operator $\hat{\rho}_N^{(1)}(t)$ of the $N$-particle system which extends the mean-field approximation through an additional collision term: 
\begin{equation}
i\hslash \, \partial_t \hat{\rho}_N^{(1)}(t) \, = \, \big[ \hat{\mathscr{H}}(t),\hat{\rho}_N^{(1)}(t) \big] + i\hat{K}(t) \, .
\label{TDHF_2}
\end{equation}
The first term on the r.h.s. describes the changes of $\hat{\rho}_N^{(1)}(t)$ as a result of the mean field $\mathscr{H}(t)$, whereas the collision term $\hat{K}(t)$, whose exact form is derived in Ref.\,\cite{gww81}, determines the effect of two-body collisions.  Due to the imaginary unit in front of the collision
term, Eq.\,(\ref{TDHF_2}) becomes time-irreversible and hence, accounts for energy dissipation. We now switch to a representation of the one-particle Hilbert space through an orthonormal set of eigenvectors $\lbrace \ket{\alpha(t)} \rbrace_{\alpha}$ of the mean field Hamiltonian $\hat{\mathscr{H}}(t)$, such that
\begin{equation}
\hat{\mathscr{H}}(t)\ket{\alpha(t)} = \epsilon_{\alpha}(t) \ket{\alpha(t)} \, .
\label{Eigenvalue_1}
\end{equation}
The single-particle density operator can then be expressed through the density matrix of this representation:
\begin{equation}
\hat{\rho}_N^{(1)}(t) \, = \, \sum_{\alpha,\beta} \ket{\alpha(t)} (\rho_N^{(1)}(t))_{\alpha,\beta} \bra{\beta(t)} \, . 
\label{Energy-Repr._1}
\end{equation}
The diagonal elements $(\rho_N^{(1)}(t))_{\alpha,\alpha} =:n(\epsilon_{\alpha},t)$ of the density matrix can be interpreted as the probability to find the particle in a state with energy $\epsilon_{\alpha}$. As in fermionic systems each state can be occupied by one particle at most, $n(\epsilon_{\alpha},t)$ is equivalent to the mean occupation number of the state $\ket{\alpha(t)}$. It is shown in Ref.\,\cite{gww81} that inserting Eq.\,(\ref{Energy-Repr._1}) into Eq.\,(\ref{TDHF_2}) and neglecting the off-diagonal elements of the density matrix leads to a master equation for the diagonal elements $n_{\alpha} \equiv n(\epsilon_{\alpha},t)$:
\begin{align}
\label{Mast.-Equ._1}
&\partial_t n_{\alpha} = \sum_{\beta,\gamma,\delta} \braket{V^2_{\alpha\beta,\gamma\delta}} \, G_E \, \big[ (1-n_{\alpha})(1-n_{\beta})n_{\gamma}n_{\delta} \, \\ \nonumber
&\qquad\qquad\qquad\qquad- \, (1-n_{\delta})(1-n_{\gamma})n_{\beta}n_{\alpha} \big] \, .
\end{align}  
Here, $\braket{V^2_{\alpha\beta,\gamma\delta}}$ denotes the second moment of the residual interaction defined through the random-matrix model and expresses the strength of pairwise interactions between the particles. The energy conserving function is $G_E \equiv G_E(\epsilon_{\alpha}+\epsilon_{\beta}, \epsilon_{\gamma}+\epsilon_{\delta})$, which unlike the delta-function has a width greater than zero for finite systems, such that collisions between particles whose single-particle states lie apart in energy space become possible. Its exact form is derived in Ref.\,\cite{gww81}.  

In order to simplify Eq.\,(\ref{Mast.-Equ._1}), we adopt an approach presented in Ref.\,\cite{gw82}, which aims to transform the equation into a partial differential equation. First, we define the transition probabilities
\begin{align}
&W_{\gamma\rightarrow\alpha} \, := \, \sum_{\beta,\delta} \, \braket{V^2_{\alpha\beta,\gamma\delta}} \, G_E \, (1-n_{\beta}) \, n_{\delta} \, ,
\label{Trans.-Prob._1}\\
&W_{\alpha\rightarrow\gamma} \, := \, \sum_{\beta,\delta} \, \braket{V^2_{\alpha\beta,\gamma\delta}} \, G_E \, (1-n_{\delta}) \, n_{\beta} \, , 
\label{Trans.-Prob._2}
\end{align}
which allow us to split the r.h.s. of Eq.\,(\ref{Mast.-Equ._1}) into a  gain and a loss term:
\begin{equation}
\partial_t n_{\alpha} \, = \, (1-n_{\alpha}) \sum_{\gamma} W_{\gamma\rightarrow\alpha} \, n_{\gamma} \; - \; n_{\alpha} \sum_{\gamma} W_{\alpha\rightarrow\gamma} \, (1-n_{\gamma}) \, . 
\label{Mast.-Equ._2}
\end{equation}
The $\gamma$-summation is then replaced by an integration, thereby introducing the densities of states $g_{\alpha} \equiv g(\epsilon_{\alpha})$ and $g_{\gamma} \equiv g(\epsilon_{\gamma})$, and substituting $W_{\gamma\rightarrow\alpha} \rightarrow W_{\gamma,\alpha} \, g_{\alpha}$ and $W_{\alpha\rightarrow\gamma} \rightarrow W_{\alpha,\gamma} \, g_{\gamma}$. Because fermions are interchangeable particles, we have $W_{\alpha,\gamma} = W_{\gamma,\alpha} \equiv W(\epsilon_{\alpha},\epsilon_{\gamma},t)$, yielding  
\begin{equation}
\partial_t n_{\alpha}  =  \int_{0}^{\infty} W_{\alpha,\gamma} \, \big[g_{\alpha} \, (1-n_{\alpha}) \, n_{\gamma} - g_{\gamma} \, (1-n_{\gamma}) \, n_{\alpha}\big] \; d\epsilon_{\gamma} \; .
\label{Mast.-Equ._3}
\end{equation}
If $\epsilon_{\alpha}$ and $t$ are fixed, $W_{\alpha,\gamma} \equiv W(\epsilon_{\alpha},\epsilon_{\gamma},t)$ is peaked around $\epsilon_{\gamma} = \epsilon_{\alpha}$ due to the finite width of $G_E$ (see Refs.\,\cite{gww81} and \cite{gw82}), so that an approximation of Eq.\,(\ref{Mast.-Equ._3}) can be obtained by a Taylor expansion of $n_{\gamma}$ and $g_{\gamma}(1-n_{\gamma})$ around $\epsilon_{\gamma} = \epsilon_{\alpha}$ to second order. 
Neglecting higher-order terms in this expansion is analogous to derivations of the (linear) Fokker-Planck equation from Pauli's master equation.
With the help of the  transport coefficients
\begin{align}
&D \, \equiv \, D(\epsilon_{\alpha},t) \, := \, \frac{1}{2} \, g_{\alpha} \,\int_{0}^{\infty} W_{\alpha,\gamma} \, (\epsilon_{\gamma}-\epsilon_{\alpha})^2 \, d\epsilon_{\gamma} \; , 
\label{D}\\[3pt]
&v \, \equiv \, v(\epsilon_{\alpha},t) \, := \, g_{\alpha}^{-1} \, \partial_{\epsilon_{\alpha}} \big[ g_{\alpha} \, D \big] \; ,
\label{v}
\end{align} 
Eq.\,(\ref{Mast.-Equ._3}) can be rewritten as a non-linear partial differential equation
\begin{equation}
\partial_t n_{\alpha} \, = \,  - \partial_{\epsilon_{\alpha}} \big[ v \, n_{\alpha} \, (1-n_{\alpha}) + n_{\alpha}^2 \,(\partial_{\epsilon_{\alpha}} D) \big] \, + \, \partial_{\epsilon_{\alpha}}^2 \big[ D \, n_{\alpha}\big] \, .
\label{Mast.-Equ._4}
\end{equation}
The non-linear terms in Eq.\,(\ref{Mast.-Equ._4}) highlight Pauli's exclusion principle, which prohibits additivity of solutions in order to prevent mean occupation numbers greater than one. One can see that the transport coefficients $D$ and $v$, which depend on the strength of residual two-body interactions, enforce the time evolution of the system and can be expected to yield a fast local equilibration. 

From the microscopic structure of the diffusion coefficient as given in Eq.\,(\ref{D}), $D$ is constant if 
$g_\alpha W_{\alpha, \gamma}$ and therefore the rate $W_{\gamma \rightarrow \alpha}$ of Eq.\,(\ref{Trans.-Prob._1}) in the master equation 
Eq.\,(\ref{Mast.-Equ._1}) is independent of energy. For constant $D$, any energy dependence of the drift $v$ according to Eq.\,(\ref{v}) is then due to the single-particle level density $g_\alpha$, and constant $v$ would require an exponential energy dependence of $g$. Hence, the transport coefficient functions $v, D$ certainly call for further detailed investigations starting from the microscopic structure of the transition probabilities, which is, however, beyond the scope of this work.

In a gradient expansion of $D$ and $v$, one could first consider a constant diffusion coefficient, and a drift coefficient that depends linearly on the energy. Such as model would be mathematically analogous to the Uhlenbeck-Ornstein model \cite{uhl30} that uses a linear Fokker-Planck equation. Although it is unlikely that the problem can be solved exactly also in the nonlinear case,
it will be interesting to tackle it numerically, and investigate quantitatively the influence of the gradients on the results. 

\section{Solution of the nonlinear diffusion equation}
In view of the difficulties to solve Eq.\,(\ref{Mast.-Equ._4}) with general dependences of the transport coefficients on the energy, we investigate here
its solutions for the simplified case of constant coefficients $D$ and $v$. Obviously, this is an
idealization that is mainly motivated by the possibility to find an exact solution. 
It reduces Eq.\,(\ref{Mast.-Equ._4}) to a \textit{fermionic diffusion equation} 
(with $n_{\alpha} \equiv n(\epsilon_{\alpha},t) \, \rightarrow \, n \equiv n(\epsilon,t)$):
\begin{equation}
\partial_t n \, \stackrel{!}{=} \,  - v \, \partial_{\epsilon} [n(1-n)] \, + \, D \, \partial_{\epsilon}^2 n \, .
\label{Mast.-Equ._5}
\end{equation}
Although this approximation appears to be very rough, it will be justified retroactively by its correct description of the equilibrium distribution, as well as its compatibility with conservation laws. In this section, we shall solve the above nonlinear equation exactly. 
\subsection{Relaxation Ansatz}
\begin{figure}
\begin{center}
\includegraphics[width=8.6cm]{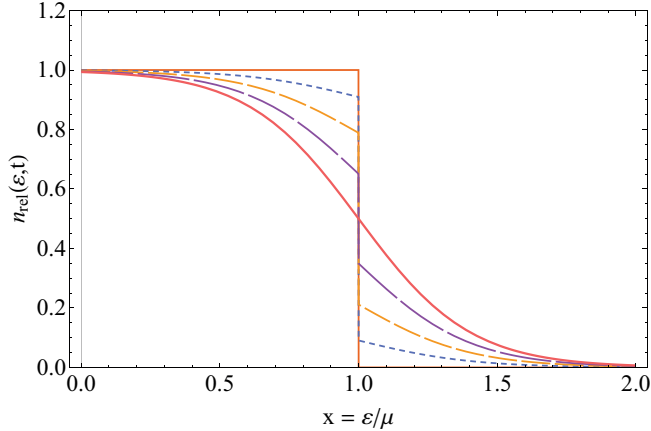}
\caption{\label{RelAns2} Approximate analytical solutions $n_\text{rel}(\epsilon,t)$ as provided by the relaxation ansatz for the initial distribution $n_0(\epsilon) = \Theta(\mu - \epsilon)$ with $\mu\beta = 5$ evaluated at different times $t/\tau_{\text{eq}}$ = 0 (solid),  0.2, 0.55, 1.2 (ordered by increasing dash length), $\infty$ (solid).}  
\end{center}
\end{figure}

Before we look for analytical solutions of Eq.\,(\ref{Mast.-Equ._5}), we examine its stationary solutions $n_{\infty}(\epsilon) \neq \text{const.}$, which solve the equation for $\partial_t n_{\infty}(\epsilon)=0$. In this case, Eq.\,(\ref{Mast.-Equ._5}) can be integrated over $\epsilon$, yielding ($n_{\infty} \equiv n_{\infty}(\epsilon)$)
\begin{equation}
\partial_{\epsilon} n_{\infty} \, \stackrel{!}{=} \, \beta \; n_{\infty} \,(n_{\infty}-1) + c \, , 
\label{Stat.-Sol._1}  
\end{equation}
where $\beta := -v/D$ and $c \in \mathbb{R}$. Under the assumption that $\lim\limits_{\epsilon \to \infty} n_{\infty} = 0 = \lim\limits_{\epsilon \to \infty} \partial_{\epsilon} n_{\infty}$, which is due to reasons of normalization, Eq.\,(\ref{Stat.-Sol._1}) is solved by the Fermi-Dirac distribution
\begin{equation}
n_{\infty}(\epsilon) \, = \, \big[1+\exp(\beta (\epsilon - \mu))\big]^{-1} \, ,
\label{Fermi_1}
\end{equation}
which correctly describes the mean occupation number of fermionic systems in thermodynamic equilibrium in the grand canonical formalism. The parameters $\beta = -v/D$ and $\mu$ from Eq.\,(\ref{Fermi_1}) can be interpreted as the inverse temperature and the chemical potential of the system, which are assumed to be fixed. Note that $\beta$ and $\mu$ are only well-defined once the $N$-particle system has reached thermodynamic equilibrium, that is after a system-specific equilibration time $\tau_{\text{eq}}$ which had been determined in Ref.\,\cite{gw82} to be $\tau_{\text{eq}}=4D/v^2$. It can be used to obtain an approximate solution of the diffusion equation Eq.\,(\ref{Mast.-Equ._5}) through the relaxation ansatz
\begin{equation}
n_{\text{rel}}(\epsilon,t) \, := \, n_{\infty}(\epsilon) + [n_0(\epsilon) - n_{\infty}(\epsilon)] \, e^{-t/\tau_{\text{eq}}} \, ,
\label{Rel.-Ans._1}
\end{equation} 
where $n_0(\epsilon)$ is an arbitrary initial distribution. Fig. \ref{RelAns2} shows $n_{\text{rel}}(\epsilon,t)$ for a simple stepped initial distribution evaluated at different times $t$. 

\subsection{Exact solution} 
Analytical solutions for physically meaningful nonlinear partial differential equations are rarely available. A prototypical exception is the Korteweg-de Vries equation \cite{kdv95}, which is of third order in the (single) spatial variable and has soliton solutions. Another example is Burgers' equation \cite{bur48}, which has the structure of a one-dimensional Navier-Stokes equation without pressure term. It has been used to describe fluid flow and, in particular, shock waves in a viscous fluid, and it can be solved through Hopf's transformation \cite{ho50}.

In case of the fermionic diffusion equation Eq.\,(\ref{Mast.-Equ._5}), one of us had proposed two analytical solution schemes in 
Ref.\,\cite{gw82}: Either start with the nonlinear transformation
\begin{equation}
n(\epsilon,t)=\frac{D}{vP(\epsilon,t)}\,\partial_\epsilon P(\epsilon,t)=\frac{D}{v}\,\partial_\epsilon \ln P(\epsilon,t)\,,
\label{nltrafo}
\end{equation}
which reduces Eq.\,(\ref{Mast.-Equ._5}) to a linear Fokker-Planck equation for $P\equiv P(\epsilon,t)$ (see below for the treatment of the missing constant)
\begin{equation}
\partial_t P=-v\,\partial_\epsilon P+D\,\partial_\epsilon^2P
\label{fpe}
\end{equation}
that is readily solvable.
Alternatively, it was proposed to perform the linear transformation
\begin{equation}
n(\epsilon,t)=\frac{1}{2v}[v-w(\epsilon,t)]\,,
\label{lintrafo}
\end{equation}
with $w\equiv w(\epsilon,t)$ obeying Burgers' equation
\begin{equation}
\partial_t w+w\partial_\epsilon w=\partial_\epsilon^2 w\,,
\label{burgers}
\end{equation}
which can then be solved analytically. In both cases, one can retransform back to $n(\epsilon,t)$ and obtain the exact solution of the nonlinear
fermionic diffusion equation Eq.\,(\ref{Mast.-Equ._5}).

Whereas in Ref.\,\cite{gw82}, and subsequently in Ref.\,\cite{gw18}, the second path was chosen to obtain the exact solution explicitly, we now make use of the first possibility to arrive at the same result. To better motivate the transformation Eq.\,(\ref{nltrafo}),
we recall that the Fermi-Dirac distribution $n_{\infty}(\epsilon)$ can be written as
\begin{equation}
n_{\infty}(\epsilon) \, = \, -\beta^{-1} \partial_{\epsilon} \ln(P_G(\epsilon)) \, ,
\label{Ansatz_1}
\end{equation}
where $P_G(\epsilon)=1+e^{-\beta(\epsilon-\mu)}$ is the grand canonical partition function of a one-state system. We extend this property to all times by making the ansatz
\begin{equation}
n(\epsilon,t) \, = \, -\beta^{-1} \partial_{\epsilon} \ln(P(\epsilon,t)) \, ,
\label{Ansatz_2}
\end{equation} 
which, inserted into Eq.\,(\ref{Mast.-Equ._5}), leads to 
\begin{align}
\partial_t  \ln{(P)} \, \stackrel{!}{=} \, &-v \, \partial_{\epsilon} \ln{(P)} \, + \, D \,\big(\partial_{\epsilon} \ln{(P)} \big)^2 \notag \\[3pt] 
&+ D \, \partial_{\epsilon}^2 \ln{(P)} \, + \, c(t) \, .
\label{Mast.-Equ._6}
\end{align}
Here, $c(t)$ is a time dependent function that arises due to an integration over $\epsilon$, but which will later turn out to be of no importance for the final expression for $n(\epsilon,t)$. After carrying out the derivatives with respect to $t$ and $\epsilon$ in Eq.\,(\ref{Mast.-Equ._6}), one arrives at the following partial differential equation for $P(\epsilon,t)$:
\begin{equation}
\partial_t P \, \stackrel{!}{=} \, -v \; \partial_{\epsilon} P + D \; \partial_{\epsilon}^2 P \, + \, c(t) \, P \, .
\label{Fokk.-Pl.-Eq.}
\end{equation}  
Apart from the last term on the r.h.s., it has the same form as the Fokker-Planck equation Eq.\,(\ref{fpe}) with constant drift coefficient $v$ and constant diffusion coefficient $D$. It can be solved by performing a Fourier transform with respect to $\epsilon$, leading to an ordinary differential equation of first order with respect to $t$:
\begin{gather}
\partial_t \tilde{P}(k,t) \, \stackrel{!}{=} \,  - [D k^2 + ivk - c(t)] \; \tilde{P}(k,t) \notag \\[3pt]
\Rightarrow \;\;\; \tilde{P}(k,t) \stackrel{!}{=} \tilde{P}_0(k) \;\; e^{-[D k^2 + ivk] \, t \, + \,C(t)} \, .
\label{Four.-Trans._3}
\end{gather}
Here, we have defined $C(t):= \int_{0}^{t} c(t') dt' + C_0$ and introduced the initial distribution in k-space; $\tilde{P}_0(k) = (2 \pi)^{-\frac{1}{2}} \int_{-\infty}^{\infty} P_0(x) e^{-ikx} dx $. The function $P(\epsilon,t)$ can be obtained by performing an inverse Fourier transform, yielding
\begin{align}
P(\epsilon,t) = \, &(4\pi Dt)^{-\nicefrac{1}{2}} \; \exp\Big(C(t)  +  \frac{t}{\tau_{\text{eq}}}  -  \frac{\beta}{2} \epsilon\Big) \;\, \times\notag \\ 
&\int_{-\infty}^{\infty} P_0 (x) \, \exp\Big(- \frac{(\epsilon-x)^2}{4Dt} + \frac{\beta}{2}x \Big) \; dx \, .
\label{P_2}
\end{align}
Here, we used $\beta = -v/D$ and $\tau_{\text{eq}}=4D/v^2$ \cite{gw82}. As the transport coefficients are related to the second moment of the residual interaction, the equilibration time and thus the speed of the equilibration process is determined by the strength of the residual interaction. 

\subsection{Generalized Partition Function}
As a consequence of Eq.\,(\ref{Ansatz_2}), the initial distribution $P_0(\epsilon)$ has to be related to the initial mean occupation number $n_0(\epsilon)$: 
\begin{gather}
n_0(\epsilon) \stackrel{!}{=} -\beta^{-1} \; \partial_{\epsilon} \ln{(P_0(\epsilon))} \notag\\
\Rightarrow \;\;\; P_0(\epsilon) = \exp \Big( -\beta \int_0^{\epsilon} n_0(y) \, dy \, + \, c' \Big)
\label{In.-Mean._Occ._1}
\end{gather}
with $c' \in \mathbb{R}$. The solution Eq.\,(\ref{P_2}) then takes the form
\begin{align}
P(\epsilon,t) \, = \, &(4\pi Dt)^{-\nicefrac{1}{2}} \; \exp\big(c' + C(t) + \tfrac{t}{\tau_{\text{eq}}}\big) \;\, \times \notag \\ 
&\exp\big(-\tfrac{\beta}{2} \epsilon\big) \;\, Z(\epsilon,t) \; ,
\label{P_4}
\end{align}
where the generalized partition function $Z(\epsilon,t)$ is defined as
\begin{gather}
Z(\epsilon,t) \, := \, \int_{-\infty}^{\infty} f(x,\epsilon,t) \, dx \;\;\;\;\; \text{with} 
\notag \\
f(x,\epsilon,t) \, := \, \exp \Big(\frac{\beta}{2} \, \Big[ x \, - \, 2 \int_0^{x} n_0(y) \; dy \Big] \, - \frac{(\epsilon-x)^2}{4Dt} \Big) \; .
\label{f}
\end{gather}
According to Eq.\,(\ref{Ansatz_2}), the mean occupation number follows as
\begin{equation}
n(\epsilon,t) \, = \, \frac{1}{2} - \beta^{-1} \partial_{\epsilon} \ln{\big(Z(\epsilon,t)\big)} \, , 
\label{n_1}
\end{equation}
or in integral representation:
\begin{equation}
n(\epsilon,t) \, = \, \frac{1}{Z(\epsilon,t)} \; \int_{-\infty}^{\infty} \Big[ \frac{1}{2} - \frac{\epsilon-x}{2vt} \Big] \, f(x,\epsilon,t) \; dx \, .
\label{n_2}
\end{equation} 
This result coincides with the result obtained in Ref.\,\cite{gw82}, where the fermionic diffusion equation Eq.\,(\ref{Mast.-Equ._5}) is solved through the second solution scheme mentioned above leading to Burgers' equation instead of to the Fokker-Planck equation\footnote{A misprint in Eq.\,(14) of Ref.\,\cite{gw82} had been corrected in Ref.\,\cite{gw18}.}. We note that the solution Eq.\,(\ref{n_1}) is independent of the terms $C(t)$ and $c'$, which emerged as integration constants. The calculation of $n(\epsilon,t)$ can be reduced to the calculation of the generalized partition function, which contains the relevant information on the time evolution and the initial distribution of the mean occupation number. In the following, this will be done for a set of simple initial distributions of the mean occupation number.

\section{Discrete-Valued Initial Conditions} 
\subsection{Step Functions}
Apart from constant distributions $n_0(\epsilon)=\text{const.}$, which can be excluded due to reasons of normalization, the simplest initial distributions of the mean occupation number are given by locally constant functions. In this case, the energy space is split up into a finite number of connected components that are accessible to the system, each of which is assigned a constant mean occupation number. Formally, we consider discrete-valued step functions of the type
\begin{equation}
n_0(\epsilon) \, =  \sum_{\substack{i=0 \\ i \, \text{even}}}^m N_i \; \Theta(\epsilon_i-\epsilon) \, \Theta(\epsilon-\epsilon_{i-1}) \; ,    
\label{Step.-Func._1}
\end{equation}
where $m\in2\mathbb{N}$, $N_i \in \mathbb{R}_{\geq 0} \; \forall i \in \lbrace 0,2,...,m \rbrace$, $\epsilon_j \in \mathbb{R}_{\geq 0} \; \forall j \in \lbrace 0,1,...,m \rbrace$ with $\epsilon_j \leq \epsilon_{j+1} \, \forall j \in \lbrace 0,1,...,m-1 \rbrace$ and $\epsilon_{-1}:= -\infty$. They correspond to occupations of the energy intervals $(\epsilon_{i-1}, \epsilon_i)$ with occupation numbers $N_i$ ($i=0,2,4,...,m$), which have to be bounded from above by one in fermionic systems. If $N_i \in \lbrace 0,1\rbrace$, the initial energy distribution of the $N$-particle-system is exactly known, whereas the case $0 < N_i < 1$ can be interpreted as a statistical distribution of an ensemble of $N$-particle systems over the intervals $(\epsilon_{i-1}, \epsilon_i)$. Let $\mathcal{S}$ be the set of all step functions of the form Eq.\,(\ref{Step.-Func._1}). Our aim is now to determine the generalized partition function $Z(\epsilon,t)$ for $n_0 \in \mathcal{S}$. For a given step function $n_0 \in \mathcal{S}$ we define the abbreviation
\begin{equation}
\delta_j := \begin{cases} 1-2N_j & j \; \text{even} \\ 1 & j \; \text{odd} \end{cases} \; ,
\label{Abbrev._1}
\end{equation} 
as well as the alternating energy moments
\begin{equation}
\mu_j \, := \, \sum_{k=0}^{j-1} \, (-1)^k \, N_{\varrho(k)} \, \epsilon_k \;\; \in \mathbb{R} \, ,
\label{Abbrev._2}
\end{equation} 
where $\varrho(k) := k + \frac{1}{2}(1-(-1)^k)$. The highest alternating energy moment $\mu_{m+1} =: \mu$ is called chemical potential. With the help of the auxiliary functions 
\begin{gather}
a_j(\epsilon,t):= \epsilon - vt \delta_j \; , 
\label{Abbrev._3} \\[3pt] 
E_j(\epsilon,t) := \beta \, \big( \frac{1}{2} \, (1+\delta_j) \, \epsilon - \mu_j \big) \, + \, (\delta^2_j -1) \,\frac{t}{\tau_{\text{eq}}} \; ,
\label{Abbrev._4}
\end{gather}
the evaluated generalized partition function can be written as
\begin{equation}
Z(\epsilon,t) = (\pi D t)^{\nicefrac{1}{2}} \, \exp \Big( \frac{t}{\tau_{\text{eq}}} - \frac{\beta}{2} \epsilon \Big) \; p(\epsilon,t) \; , 
\label{Z_simple}
\end{equation}
where we have introduced
\begin{gather}
p(\epsilon,t) := \sum_{j=0}^{m+1} e^{E_j(\epsilon,t)} \; p_j(\epsilon,t) \; ,
\label{p} \notag \\
p_j(\epsilon,t) \, := \, \text{erf}\Big( \frac{\epsilon_j - a_j(\epsilon,t)}{(4Dt)^{\nicefrac{1}{2}}} \Big) - \text{erf}\Big( \frac{\epsilon_{j-1} - a_j(\epsilon,t)}{(4Dt)^{\nicefrac{1}{2}}} \Big) \; ,
\label{p_j}
\end{gather}
and set $\epsilon_{m+1} := \infty$. Inserting Eq.\,(\ref{Z_simple}) into Eq.\,(\ref{n_1}) then yields the following solution for the mean occupation number:
\begin{equation}
n(\epsilon,t) \, = \, 1 - \beta^{-1} \partial_{\epsilon} \ln{(p(\epsilon,t))} \, .
\label{n_3}
\end{equation} 
We note that the generalized partition function $Z(\epsilon,t)$ has been replaced by the function $p(\epsilon,t)$, which differs from $Z(\epsilon,t)$ by irrelevant prefactors. 

\subsection{Stationarity Conditions}\label{Stationarity Conditions}
Eq.\,(\ref{n_3}) is valid for any step function $n_0 \in \mathcal{S}$. However, as the mean occupation number has to be normalizable for all times and especially for  $t \to \infty$, we can allow only those step functions to be considered as valid initial distributions for which the sequence $(n(\epsilon,t))_t$ converges to a non-constant stationary solution. For $N_0 \leq \frac{1}{2}$, the solution $n(\epsilon,t)$ can be shown to vanish in thermodynamic equilibrium; $\lim\limits_{t \to \infty} n(\epsilon,t) = 0$. For $N_0 > \frac{1}{2}$, its asymptotic behaviour for $t \to \infty$ is given by 
\begin{gather}
n(\epsilon,t) \; \approx 
N_0 \, \Big[1 + \exp\Big(\beta \, (N_0\epsilon - \mu_{m+1}) + \frac{t}{\tau_{\text{eq}}} (1 - \delta^2_0)\Big)\Big]^{-1} \, .
\label{Asymp}
\end{gather}
We conclude that the convergence of $n(\epsilon,t)$ towards a stationary solution only depends on the parameter $N_0$, which describes the occupation number assigned to states with negative energy. For $N_0 > \frac{1}{2}$ we can expect a stationary limit if the time dependent term in Eq.\,(\ref{Asymp}) drops out. This is the case if 
\begin{equation}
1 - \delta^2_0 \stackrel{!}{=} 0 \;\;\; \Rightarrow \;\;\; \delta_0 \equiv 1-2N_0 \stackrel{!}{=}  -1 \;\;\; \Rightarrow \;\;\; N_0 \stackrel{!}{=} 1 \; .
\label{Stat.-Cond._1} 
\end{equation}  
Therefore, step functions $n_0 \in \mathcal{S}$ converge to a stationary solution if and only if $N_0=1$. We pool all the step functions fulfilling this condition in the subset $\mathcal{S}_1 := \lbrace n_0 \in \mathcal{S} \, \vert \, N_0 =1 \rbrace \subset \mathcal{S}$. According to Eq.\,(\ref{Asymp}), the stationary limit in this case is given by
\begin{equation}
n_{\infty} (\epsilon) := \lim\limits_{t \to \infty} n(\epsilon,t) \, = \, \big[1 + \exp(\beta (\epsilon-\mu))\big]^{-1} \, ,
\label{Fermi_2}
\end{equation}
where we used the definition of the chemical potential $\mu=\mu_{m+1}$. Note that there are no formal restrictions for the remaining occupation numbers $N_j$ ($j=1,...,m$), so that initial occupations of a state with positive energy with more than one particle would also lead to the correct equilibrium distribution. However, the Pauli exclusion principle demands that we restrict all $N_j$ to the interval $[0,1]$.

\subsection{Example Solutions}
\begin{figure}
\begin{center}
\includegraphics[width=8.6cm]{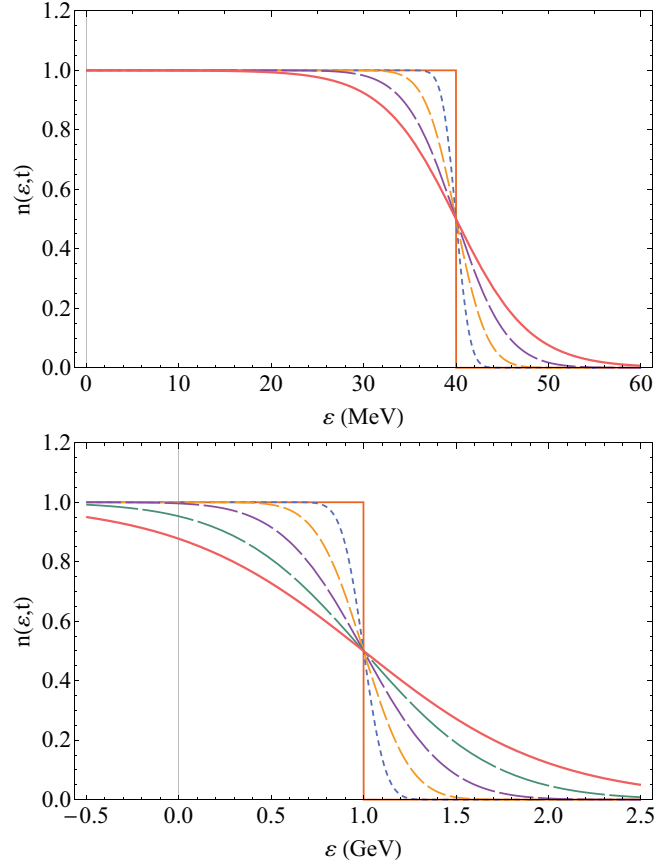}
\caption{\label{nlowhigh1}Mean occupation number $n(\epsilon,t)$ evaluated from the fermionic diffusion equation at different times for the initial distribution
$n_0(\epsilon) = \Theta(\epsilon_0 - \epsilon)$ in a low- and a high-energy case.
Upper diagram:
$\epsilon_0= 40 \, \text{MeV}$, $T = 4 \, \text{MeV}$, $\tau_{\text{eq}} = 3.2 \cdot 10^{-23} \, s$, $n(\epsilon,t)$ evaluated at 
$t/\tau_{\text{eq}}$ = 0, 0.013, 0.078, 0.313, $\infty$.
Lower diagram:
 $\epsilon_0= 1 \, \text{GeV}$, $T = 510 \, \text{MeV}$, $\tau_{\text{eq}} = 0.\bar{3} \cdot 10^{-23} \, \text{s}$, $n(\epsilon,t)$ evaluated at 
$t/\tau_{\text{eq}}=$  0, 0.005, 0.024, 0.09, 0.33, $\infty$ (ordered by increasing dash length).}  
\end{center}
\end{figure}

The transport coefficients $D$ and $v$ have been introduced as a result of a microscopic theory for the $N$-particle system. However, they could be connected to macroscopic variables such as the temperature $T := \beta^{-1} \equiv -D/v$ and the equilibration time $\tau_{\text{eq}} \equiv 4D/v^2$. If we assume these as given, we can therefore determine the microscopic transport coefficients as
\begin{equation}
D = \frac{4}{\tau_{\text{eq}}} \, T^2 \, , \;\;\;\;\;\;\;\;\; v = -\frac{4}{\tau_{\text{eq}}} \, T \, .
\label{Transp.-Coef._1}
\end{equation}
In the following, some example solutions for initial distributions $n_0 \in \mathcal{S}_1$ are discussed. In the simplest case $n_0(\epsilon) = \Theta(\epsilon_0 - \epsilon)$, the solution Eq.\,(\ref{n_3}) takes the form
\begin{equation}
n(\epsilon,t) \, = \, 1 - \beta^{-1} \; \partial_{\epsilon} \ln\big( p_0 + e^{E_1}p_1 \big) \, ,
\label{n_4}
\end{equation} 
where
\begin{align}
&p_0 \, \equiv \, p_0(\epsilon,t) \, = \, 1 + \text{erf}\Big( \frac{\epsilon_0 - \epsilon - vt}{(4Dt)^{\nicefrac{1}{2}}} \Big) \, , \\
&p_1 \, \equiv \, p_1(\epsilon,t) \, = \, 1 - \text{erf}\Big( \frac{\epsilon_0 - \epsilon + vt}{(4Dt)^{\nicefrac{1}{2}}} \Big) \, , \\ 
&E_1 \, \equiv \, E_1(\epsilon,t) \, = \, \beta (\epsilon - \epsilon_0) \, .
\end{align}
The initial maximum energy $\epsilon_0$ corresponds to the chemical potential $\mu$ in equilibrium. Fig. \ref{nlowhigh1} shows the solution Eq.\,(\ref{n_4}) for a low-energy case with temperature $T=4\, \text{MeV}$ and equilibration time $\tau_{\text{eq}} = 3.2 \cdot 10^{-23} \, \text{s}$ as well as for a high-energy case with temperature $T = 510\, \text{MeV}$ and equilibration time $\tau_{\text{eq}} = 0.\bar{3} \cdot 10^{-23} \, \text{s}$. 
One can see that the distributions approach the expected Fermi-Dirac distribution for $t \to \infty$ and are point-symmetric around $\epsilon=\epsilon_0$ at all times: 
\begin{equation}
n(\mu + \epsilon,t)=1-n(\mu-\epsilon,t) \;\;\;\;\; \forall \, \epsilon \in \mathbb{R}, \, t \in \mathbb{R}_{\geq 0} \; .
\label{Point-Symm._1}
\end{equation}
In contrast to the relaxation ansatz shown in Fig. \ref{RelAns2}, the equilibration takes place comparatively fast and the distributions vary smoothly for $t\in \mathbb{R}_{>0}$ with no discontinuities at $\epsilon=\epsilon_0$. Whereas in the low-energy case, all states on the negative real axis are fully occupied, occupation numbers below one occur for $\epsilon < 0$ in the high-energy case. This circumstance will later be interpreted as the creation of antiparticles. 

\begin{figure}
\begin{center}
\includegraphics[width=8.6cm]{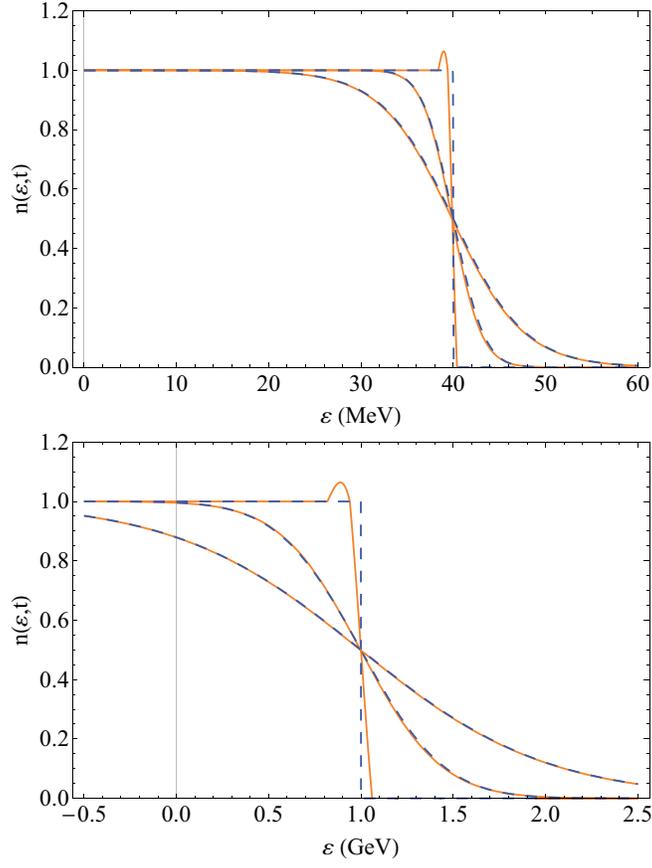}
\caption{\label{num_low+high_1}Comparison of the numerical (solid) and analytical (dashed) solutions for $n(\epsilon,t)$ evaluated at different times for the initial distribution $n_0(\epsilon) = \Theta(\epsilon_0 - \epsilon)$ in a low- and a high-energy case.
Upper diagram:
$\epsilon_0= 40 \, \text{MeV}$, $T = 4 \, \text{MeV}$, $\tau_{\text{eq}} = 3.2 \cdot 10^{-23} \, \text{s}$, $n(\epsilon,t)$ evaluated at 
$t/\tau_{\text{eq}}$ = 0, 0.078, $\infty$. 
Lower diagram:
 $\epsilon_0= 1 \, \text{GeV}$, $T = 510 \, \text{MeV}$, $\tau_{\text{eq}} = 0.\bar{3} \cdot 10^{-23} \, \text{s}$, $n(\epsilon,t)$ evaluated at 
$t/\tau_{\text{eq}}=$  0, 0.09, $\infty$.}
\end{center}
\end{figure}

In Fig. \ref{num_low+high_1} the analytical solutions for the low- and high-energy case are compared to the respective numerical solutions of Eq.\,(\ref{Mast.-Equ._5}) obtained with the \textit{NDSolve} routine of \textit{Mathematica 11.1}. Apart from the initial case at $t=0$, in which the numerical solutions are smoothed out by a hump around the Fermi edge in order to guarantee differentiability, the numerical solutions (solid) coincide with the analytical solutions (dashed).

Regarding the high-energy case, the solutions of our fermionic diffusion equation may turn out to be relevant for the description of the local equilibration of quarks in heavy-ion collisions at relativistic energies, such as PbPb collisions at energies reached at the Large Hadron Collider (LHC). As an example, at a centre-of-mass energy of 5.02 TeV per particle pair, the initial central temperature $T$ is above 500 MeV \cite{hnw17}, see the calculations shown in 
Figs.\,\ref{nlowhigh1} and 3, bottom frames. 

Since our present work is dealing with fermions only, it refers, in particular, to the local equilibration of valence quarks. These reside in the fragmentation distributions \cite{gw16}, not in the anisotropic fireball source which harbours low-$x$ gluons. In the fragmentation regions, the anisotropy is not as pronounced as in the fireball source, such that the isotropy assumption that is implicit in our analytical solution is probably reasonable. Moreover, to treat the problem of local equilibration of fermions, one may replace the energy variable $\epsilon$ by the transverse energy $\epsilon_\perp$ since the question of local equilibration can also be studied in the transverse degrees of freedom alone. In the transverse plane isotropy is, of course, fulfilled in central collisions. 

\begin{figure}
\begin{center}
\includegraphics[width=8.6cm]{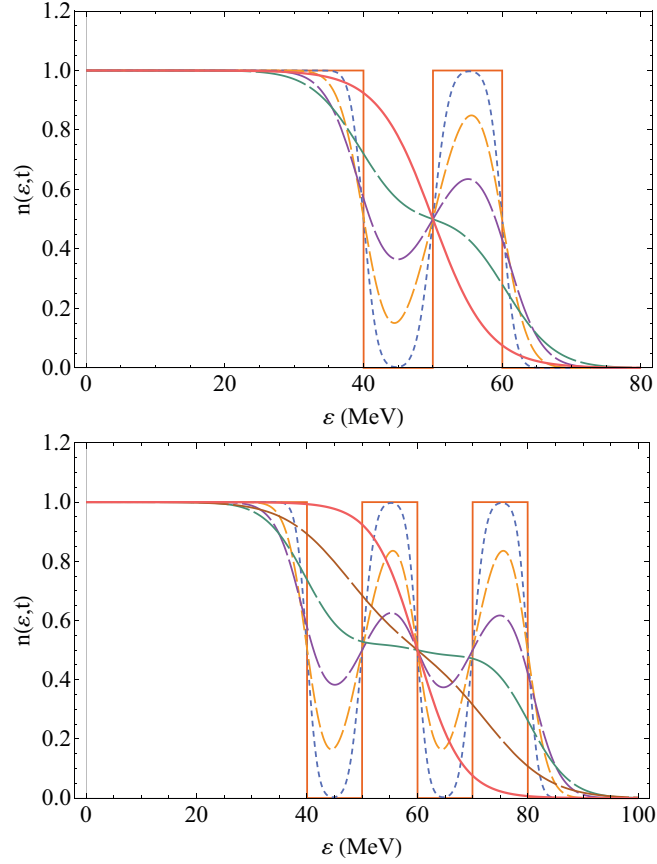}
\caption{\label{nlowhigh2}Mean occupation number $n(\epsilon,t)$ evaluated at different times for different initial distributions in the low-energy case ($T = 4 \, \text{MeV}$, $\tau_{\text{eq}} = 3.2 \cdot 10^{-23} \, \text{s}$).
Upper diagram:
$n_0(\epsilon)= \Theta(\epsilon_0 - \epsilon) +  \Theta(\epsilon - \epsilon_1)  \Theta(\epsilon_2 - \epsilon) , \epsilon_0$ = 40 MeV, $\epsilon_1 $= 50 MeV, $\epsilon_2$ = 60 MeV, $n(\epsilon,t)$ evaluated at 
$t/\tau_{\text{eq}}$ = 0, 0.021, 0.094, 0.234, 0.594, $\infty$. 
Lower diagram:
$n_0(\epsilon)= \Theta(\epsilon_0 - \epsilon) +  \Theta(\epsilon - \epsilon_1)  \Theta(\epsilon_2 - \epsilon) +  \Theta(\epsilon - \epsilon_3)  \Theta(\epsilon_4 - \epsilon), \epsilon_0$ = 40 MeV, $\epsilon_1 $= 50 MeV, $\epsilon_2$ = 60 MeV, $\epsilon_3 $= 70 MeV, $\epsilon_4$ = 80 MeV, $n(\epsilon,t)$
evaluated at $t/\tau_{\text{eq}}$ = 0, 0.022, 0.1, 0.253, 0.656, 1.594, $\infty$.} 
\end{center}
\end{figure}

Fig. \ref{nlowhigh2} shows the evolution of the mean occupation number for different initial distributions $n_0 \in \mathcal{S}_1$ in the low-energy case.
In both cases, the distributions are bounded from above by one, $n(\epsilon,t) \leq 1 \;\; \forall \, \epsilon \in \mathbb{R}, \, t \in \mathbb{R}_{\geq 0}$, such that Pauli's exclusion principle is respected. The initial distributions contain unoccupied gaps below and occupied bands above the chemical potential $\mu$, which are filled and emptied respectively during the equilibration process. Due to a suitable choice of initial conditions, all the solutions portrayed in Fig. \ref{nlowhigh2} are point-symmetric around $\epsilon = \mu$ in the sense of Eq.\,(\ref{Point-Symm._1}), which is not generally the case for an arbitrary initial distribution $n_0 \in \mathcal{S}_1$.

\section{Conservation Laws}  

\subsection{Particle-Number Conservation} 
\begin{figure}
\begin{center}
\includegraphics[width=7.4cm]{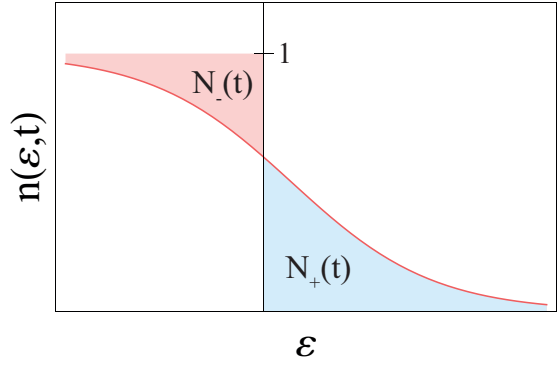}
\caption{\label{fig5} Sketch of particle-antiparticle creation at relativistic energies. The Dirac sea in the negative-energy domain is occupied and prevents particles ($N_+$) from dropping below zero energy. Holes in the Dirac sea are interpreted as antiparticles ($N_-$). The number of particles minus antiparticles $N=N_+(t) - N_-(t)$ is preserved by the nonlinear diffusion equation.}  
\end{center}
\end{figure}

As we have seen in section \ref{Stationarity Conditions}, solutions for the mean occupation number with discrete-valued initial distributions can only be expected to have stationary limits for $t \to \infty$ if $N_0=1$. Because of Pauli's exclusion principle $n(\epsilon,t) \leq 1 \; \forall \, \epsilon,t$, this corresponds to a full occupation of all states with negative energy. In the framework of the theoretical model of the Dirac sea (see Ref.\,\cite{pamd30}), holes (not fully occupied states) in the Dirac sea are interpreted as antiparticles. We therefore define the number of particles $N_+$ and antiparticles $N_-$ in the system as \cite{gw18} (see Fig. \ref{fig5})
\begin{align}
&N_+(t) \, := \, \int_{0}^{\infty} n(\epsilon,t) \, g(\epsilon) \, d\epsilon \, , 
\label{Part.-Numb._1}\\[3pt]
&N_-(t) \, := \, \int_{-\infty}^{0} [1-n(\epsilon,t)] \, g(\epsilon) \, d\epsilon \, .
\label{Ant.-Part.-Numb._1}
\end{align}

\begin{figure}
\begin{center}
\includegraphics[width=8.6cm]{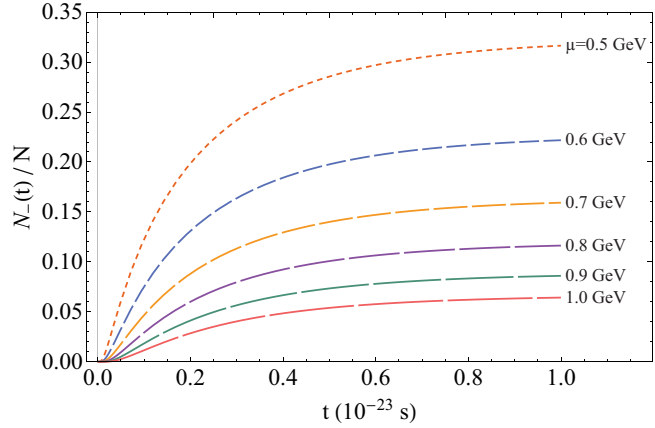}
\caption{\label{t_high_1} Time evolution of the relative antiparticle number $N_-(t)/N$ with the initial distribution $n_0(\epsilon) = \Theta(\mu-\epsilon)$ for six different chemical potentials $\mu = 0.5 - 1$ GeV in the high-energy case ($T = 510$ MeV,  $\tau_{\text{eq}}  = 0.33\cdot 10^{-23}$ s). }  
\end{center}
\end{figure}

\noindent
Here, we have introduced the density of states $g(\epsilon)$ which for simplicity is assumed to be constant; $g(\epsilon) \equiv g \in \mathbb{R}$. The condition $N_0\stackrel{!}{=} 1$ then implies the absence of antiparticles at initial time $t=0$. The integrals in Eqs.\,(\ref{Part.-Numb._1}) and (\ref{Ant.-Part.-Numb._1}) can be solved by plugging in the solution Eq.\,(\ref{n_3}) for $n(\epsilon,t)$, yielding
\begin{align}
&N_+(t) \; = \; g \, \big[\; \mu + \beta^{-1} \ln\big( \tfrac{1}{2} \, p(0,t)\big) \big] \; ,
\label{Part.-Numb._3}\\[3pt]
&N_-(t) \; = \; g \, \beta^{-1} \ln\big( \tfrac{1}{2} \, p(0,t)\big) \; .   
\label{Ant.-Part.-Numb._3}
\end{align}
We immediately see that $N_+(t) \equiv g\mu + N_-(t)$, such that the effective particle number $N$ is conserved over time:
\begin{equation}
N \, = \, N_+(t) - N_-(t) \, = \, g\mu \, \equiv \, \text{const.}
\label{EffPartNumb1}
\end{equation}  
Particles and antiparticles are produced with the same rate $R(t):=\partial_t N_+(t) \equiv \partial_t N_-(t)$, so that for every produced particle an antiparticle is produced to keep the effective particle number constant. With Eq.\,(\ref{EffPartNumb1}) we can identify $g=N/ \mu$ and define the relative particle and antiparticle numbers
\begin{align}
&n_+(t) \, := \,  N_+(t) / N \; \equiv \; 1 + (\mu\beta)^{-1} \ln\big( \tfrac{1}{2} \, p(0,t)\big) \, , \\[3pt]
&n_-(t) \, := \,  N_-(t) / N \; \equiv \; (\mu\beta)^{-1} \ln\big( \tfrac{1}{2} \, p(0,t)\big) \, .    
\end{align}
Fig. \ref{t_high_1} shows the evolution of the relative antiparticle number $n_-(t)$ for the initial distribution $n_0(\epsilon) = \Theta(\mu-\epsilon)$ with different chemical potentials $\mu$ in the high-energy case. One can see that in all cases the relative antiparticle number strives towards a limit value for $t \to \infty$, which is given by   
\begin{equation}
\lim\limits_{t \to \infty}  n_-(t) \; = \; (\mu\beta)^{-1} \, \ln (1+e^{-\mu\beta}) \; .
\label{Lim._n-}
\end{equation}
It only depends on the ratio $\mu\beta$ of the chemical potential and the temperature of the system and vanishes for $\mu\beta \to \infty$.  

\subsection{Energy Conservation}  
Both particles and antiparticles contribute to the total internal energy of the system. Based on the expressions Eq.\,(\ref{Part.-Numb._1}) and Eq.\,(\ref{Ant.-Part.-Numb._1}) for the particle and antiparticle number, we define the respective energy portions according to 
\begin{align}
&E_+(t) \, := \, \int_{0}^{\infty} \epsilon \; n(\epsilon,t) \, g \, d\epsilon \, , 
\label{Part.-En._1}\\[3pt]
&E_-(t) \, := \, \int_{-\infty}^{0} (-\epsilon) \, [1-n(\epsilon,t)] \, g \, d\epsilon \, .
\label{Ant.-Part.-En._2}
\end{align}
\begin{figure}
\begin{center}
\includegraphics[width=8.6cm]{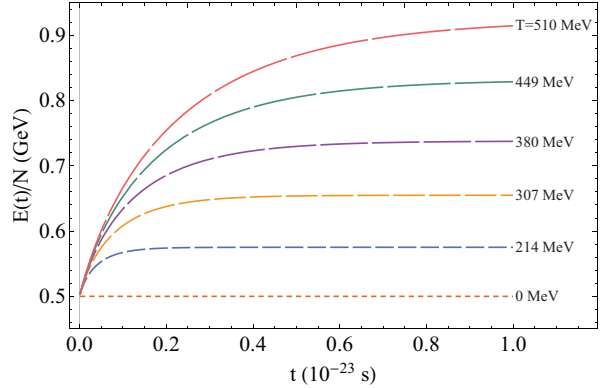}
\caption{\label{e_high_1}Time evolution of the relative internal energy $E(t)/N$ for the initial distribution $n_0(\epsilon) = \Theta(\mu-\epsilon)$ with $\mu = 1$ GeV for six different temperatures $T = 0 - 510$ MeV, while the diffusion coefficient $D = 3.1\cdot 10^{23} \text{GeV}^2 \text{s}^{-1}$ is held constant.}
\end{center}
\end{figure}
Again we adopt the constant density of states $g=N/\mu$ from before to simplify the analytical calculations. The minus sign in Eq.\,(\ref{Ant.-Part.-En._2}) represents the fact that, although holes in the Dirac sea are located on the negative energy domain, the corresponding antiparticles are interpreted as real particles with positive energy. The total internal energy of the system is given by
\begin{equation}
E(t) \, := \, E_+(t) + E_-(t) \, .
\label{Tot.-En._1}
\end{equation}
It includes the kinetic and potential energy of the particles and antiparticles as well as the thermal energy. The latter is converted from the kinetic (or relativistic) energy of the collision partners in two-body collisions. However, as the temperature is only a well-defined variable once the $N$-particle system has reached thermodynamic equilibrium, $E(t)$ has to increase over time to include the additional thermal energy next to the initial energy content $E(0)$, which is determined solely by the initial distribution $n_0 \in \mathcal{S}_1$ and is thus completely independent of the temperature $T$. In order to compare the initial energy content with the total internal energy in thermodynamic equilibrium, we evaluate $\lim\limits_{t \to \infty} E(t)$ for an arbitrary initial distribution $n_0 \in \mathcal{S}_1$ with chemical potential $\mu$: 
\begin{align}
\lim\limits_{t \to \infty} E_+(t) \; &= \; g \int_{0}^{\infty} \epsilon \, \big[1+\exp(\beta(\epsilon-\mu))\big]^{-1} d\epsilon 
= -\frac{g}{\beta^2} \, \text{Li}_2(- e^{\beta \mu}) \; , \\[3pt] 
\lim\limits_{t \to \infty} E_-(t) \; &= \; - g \int_{-\infty}^{0} \epsilon \, \big[1+\exp(\beta(\mu - \epsilon))\big]^{-1} d\epsilon 
= -\frac{g}{\beta^2} \text{Li}_2(- e^{- \beta \mu}) \; , \\[3pt]
\Rightarrow \;\; \lim\limits_{t \to \infty} E(t) \; &= \; -\frac{g}{\beta^2} \, \big[ \text{Li}_2(- e^{\beta \mu}) + \text{Li}_2(- e^{- \beta \mu})  \big] 
\equiv \; N\, \big(\frac{1}{2} \, \mu + \frac{\pi^2}{6\mu\beta^2} \big) \, .
\label{Tot.-En._2}
\end{align}
Here, we have made use of the complex dilogarithm
\begin{equation}
\text{Li}_2 (z) \, := \, \sum_{k=1}^{\infty} \frac{z^k}{k^2} \, , \;\;\; z \in \mathbb{C} \, ,
\label{Dilog.}
\end{equation}
and applied the addition theorem $\text{Li}_2 (-z) + \text{Li}_2 (-z^{-1}) = \frac{1}{2} \ln^2(z) + \frac{\pi^2}{6}$, $z \in \mathbb{C}$. We compare the result Eq.\,(\ref{Tot.-En._2}) to the initial energy content of the simple initial distribution $n_0(\epsilon) = \Theta(\epsilon_0 - \epsilon)$, which is given by  
\begin{equation}
E(0) \; = \; g \, \int_{0}^{\infty} \epsilon \, \Theta(\epsilon_0 - \epsilon) \, d\epsilon \; = \; \frac{N}{2} \, \mu \; ,  
\end{equation} 
using $\epsilon_0 = \mu$. The thermal energy is given by the difference between the final and initial total internal energy:
\begin{equation}
E_{\text{th}}(T) \; := \; \lim\limits_{t \to \infty} E(t) - E(0) \; = \; N\, \frac{\pi^2}{6\mu} \, T^2 \; .  
\end{equation}
This result coincides with the expression for the internal energy of an ideal Fermi gas for the case $\mu\beta \gg 1$ as obtained from Sommerfeld's expansion, given that the density of states is set constant.
Fig. \ref{e_high_1} shows the evolution of the relative internal energy $E(t)/N$ for different temperatures in the simple case  $n_0(\epsilon) = \Theta(\epsilon_0 - \epsilon)$ with $\epsilon_0 = 1\, \text{GeV}$.

One can see that in all cases the curves are monotonously increasing over time and strive towards the limit value given in Eq.\,(\ref{Tot.-En._2}), which grows quadratically with temperature $T$. The higher the temperature, the longer it takes for the relative internal energy to reach its equilibrium value. Only for $T=0$ the relative internal energy is independent of time, such that the initial relative energy content corresponds to the relative internal energy in equilibrium. In the high-energy case, where $T =$ 510 MeV, the initial relative energy content $E(0)/N=$ 500 MeV nearly doubles over time, reaching a final value of $\lim\limits_{t \to \infty} \, E(t)/N$ = 928 MeV. The thermal energy -- which is taken from the kinetic, or relativistic energy of the system -- can therefore make a significant contribution to the total internal energy of the system.

\section{Conclusion and outlook}
To summarize, we have schematically  modelled the time evolution of an equilibrating finite fermionic system through a non-linear partial differential fermionic diffusion equation, which resulted from a suitable transformation of the master equation for the mean occupation number \cite{gw82}. In the limit of constant transport coefficients $D$ and $v$, the equation could be solved analytically, yielding the correct limit value behaviour in thermodynamic equilibrium. Neglecting the energy and time dependence of the transport coefficients represents a rough but important approximation, which we have justified by the equilibrium properties and the analytical solvability of the resulting differential equation. 

Taking into account the dependences of $D$ and $v$ on the energy would require a detailed understanding of the energy conserving function and the microscopic interactions between the particles, and would lead to a highly nonlinear diffusion equation which can only be solved numerically. It would therefore become difficult to link the transport coefficients and thus the strength of residual interactions to macroscopic variables such as the temperature and the equilibration time. A microscopic calculation of the transport coefficients
from Eqs.\,(\ref{D}) and (\ref{v}) is, however, desirable.

The analytical solutions of the fermionic diffusion equation with constant coefficients were evaluated for a set of simple discrete-valued initial distributions, which were shown to converge towards a Fermi-Dirac distribution if and only if the Dirac sea is fully occupied at initial time $t=0$. More general initial conditions describing statistical distributions over the single-particle states are conceivable, yet have to be checked for the correct limit-value behaviour in thermodynamic equilibrium. It appears that the full occupation of the Dirac sea at initial time $t=0$ represents a necessary condition. Under the assumption of a constant density of states, the solutions preserve the effective particle number over time if the creation of antiparticles is taken into account. 

More realistic densities of states display an energy dependence, such as $g(\epsilon) \propto \sqrt{\epsilon}$ in non-relativistic or $g(\epsilon) \propto \epsilon^2$ in relativistic particle dynamics. Introducing non-constant densities of states leads, however, to a violation of the conservation of the effective particle number as defined in Eq.\,(\ref{EffPartNumb1}). Hence, the chemical potential $\mu$ must then be renormalized to still secure particle-number conservation.

The total internal energy of the system was shown to gain additional thermal energy over time, which coincides with the expression for the energy of an ideal Fermi gas in the limit $\mu\beta \gg 1$ if antiparticles are interpreted as real particles with positive energy. The additional thermal energy is taken from the kinetic energy of the collision partners, at very high energies from the available relativistic energy.

The analytical model investigated in this work has been built on the quantum mechanical description of fermionic systems through density operators consisting of antisymmetric states. An analogous approach can be made for the description of equilibrating bosonic systems. This has been done in Ref.\,\cite{gw18}, where a bosonic diffusion equation for the mean occupation number $n \equiv n(\epsilon,t)$ of the single-particle energy-states is obtained similarly to the fermionic case:
\begin{equation}
\partial_t n \, = \,  - v \, \partial_{\epsilon} [n(1+n)] \, + \, D \, \partial_{\epsilon}^2 n \, .
\label{Bos.-Diff.-Equ._1}
\end{equation}
One can see that the only difference to the fermionic diffusion equation lies in a plus sign instead of a minus sign in the nonlinear term on the r.h.s. of the equation. The stationary solution of Eq.\,(\ref{Bos.-Diff.-Equ._1}) is given by the Bose-Einstein distribution 
\begin{equation}
n_{\infty}(\epsilon) \, = \, \big[\exp(\beta (\epsilon - \mu))-1\big]^{-1} \, 
\end{equation}
with $\beta := - v/D$ and $\mu \in \mathbb{R}$, which correctly describes bosonic systems in thermodynamic equilibrium. Analytical solutions of Eq.\,(\ref{Bos.-Diff.-Equ._1}) have been obtained in Ref.\,\cite{gw18} analogously to the fermionic case. 
  
However, the analytical modelling of the equilibration process through Eq.\,(\ref{Bos.-Diff.-Equ._1}) is more challenging than in the fermionic case, because Bose-Einstein condensation may occur at sufficiently low temperatures \cite{dal99}, leading to a final state that differs from the purely thermal distribution \cite{svi91}. The buildup of the thermal tail for bosons in the ultraviolet, as well as the population of the condensate that is accounted for indirectly through the conservation of the total particle number in Eq.\,(\ref{Bos.-Diff.-Equ._1}) have been examined in Ref.\,\cite{gw18}, and applied to cold quantum gases in Ref.\,\cite{gw18a}.

\bigskip
\noindent
\bf{Acknowledgements}
\rm

TB is now at the University of Cambridge, UK.
We thank Matthias Bartelmann for discussions and remarks.


\bibliographystyle{elsarticle-num}
\bibliography{gw_18}
\end{document}